\documentclass{article}
\usepackage{iclr2027_conference,times}
\usepackage{amsmath,amsfonts,bm}

\def\eqref#1{equation~\ref{#1}}
\def\1{\bm{1}}

\DeclareMathAlphabet{\mathsfit}{\encodingdefault}{\sfdefault}{m}{sl}
\SetMathAlphabet{\mathsfit}{bold}{\encodingdefault}{\sfdefault}{bx}{n}

\usepackage{amsmath,amssymb}
\usepackage{booktabs}
\usepackage{graphicx}
\usepackage{multirow}
\usepackage{float}
\usepackage{capt-of}
\usepackage{microtype}
\usepackage{xcolor}
\usepackage{colortbl}
\usepackage{hyperref}
\usepackage{url}
\usepackage{wrapfig}
\hypersetup{hidelinks}

\newcommand{\method}{SpikCTU}

\newcommand{\T}{\mathsf{T}}

\title{SpikCTU: A Compact Spiking Network with Adaptive Recurrent Computation for Speech Command Recognition}
\author{
Jin Li$^{1}$ \quad
Wenbin Jiang$^{1}$\thanks{Corresponding author: \href{mailto:wbjiang@hdu.edu.cn}{\texttt{wbjiang@hdu.edu.cn}}.} \quad
Tao Chen$^{2}$ \quad
Fei Wen$^{3}$\\[2pt]
$^{1}$Hangzhou Dianzi University\\
$^{2}$Olympus Hedge Fund Investment Co., Ltd.\\
$^{3}$Shanghai Jiao Tong University
}

\makeatletter
\AtBeginDocument{
\setlength{\abovedisplayskip}{4pt plus1pt minus2pt}
\setlength{\belowdisplayskip}{4pt plus1pt minus2pt}
\setlength{\abovedisplayshortskip}{2pt plus0.5pt minus1pt}
\setlength{\belowdisplayshortskip}{2pt plus0.5pt minus1pt}

\renewcommand\section{\@startsection{section}{1}{\z@}
  {-1.2ex plus -0.3ex minus -0.2ex}
  {0.6ex plus 0.2ex minus 0.1ex}
  {\large\sc\raggedright}}
\renewcommand\subsection{\@startsection{subsection}{2}{\z@}
  {-1.0ex plus -0.3ex minus -0.2ex}
  {0.4ex plus 0.2ex minus 0.1ex}
  {\normalsize\sc\raggedright}}
\renewcommand\subsubsection{\@startsection{subsubsection}{3}{\z@}
  {-0.8ex plus -0.2ex minus -0.1ex}
  {0.3ex plus 0.1ex minus 0.1ex}
  {\normalsize\sc\raggedright}}

\setlength{\topsep}{2pt plus1pt minus1pt}
\setlength{\itemsep}{1pt plus0.5pt minus0.5pt}
\setlength{\parsep}{1pt plus0.5pt minus0.5pt}
}
\makeatother

\iclrfinalcopy
\begin{document}
\maketitle
\fancyhead{}

\begin{abstract}
Spiking Neural Networks offer event-driven, energy-efficient computation for speech command recognition, yet most existing methods follow fixed computation paths, allocating redundant computation to easy samples while providing limited insight into how evidence evolves during inference. 
To address these challenges, we introduce \textbf{\method}, a compact spiking network with adaptive recurrent computation for speech command recognition. Specifically, a hybrid spiking embedding module preserves continuous input information alongside spike features to construct a fixed temporal memory. A shared Continuous Thinking Unit (CTU) repeatedly reads this memory using keys derived from spikes and continuous values, and updates a binary recurrent state with lightweight stage adapters. Furthermore, combining a fixed base prediction with accumulated corrections at each step enables learned aggregation over all steps or confidence-based early exit using only the predictions available so far.
Experiments on three speech command recognition benchmarks show that \method\ achieves competitive accuracy with only 0.16--0.17\,M parameters, and the early exit strategy can substantially further reduce computational cost without compromising accuracy. In addition, trajectory and attention analyses reveal that recurrent steps enable error correction and progressively refine predictions. Code is available at \url{https://github.com/Rivflyyy/spikctu}.

\end{abstract}

\section{Introduction}
\label{sec:introduction}
Spiking Neural Networks (SNNs), widely regarded as the third generation of neural networks~\citep{maass1997networks}, communicate through biologically plausible, sparse spike signals. Spike-driven synaptic operations can replace multiply--accumulate (MAC) operations in artificial neural networks (ANNs) with accumulations (ACs), potentially reducing energy consumption~\citep{nunes2022spiking,hu2024largescale,sabbella2025promise}. This energy advantage is further amplified on neuromorphic hardware, where sparse, event-driven spikes map directly onto low-energy asynchronous circuits~\citep{bos2024micropower,wang2025background}. In recent years, spiking models have achieved competitive performance across vision and event-based benchmarks and have begun to match ANN accuracy on speech command recognition (SCR)~\citep{wang2024efficient,li2026lightweight,wang2026spikcommander}. However, the realized efficiency depends on how inference is configured: fixed timesteps, forward passes that evaluate every step, and dense spike propagation can waste computation on easy inputs, motivating conditional computation that allocates compute according to input difficulty.

\begin{figure}[t]
  \centering
  \includegraphics[width=\linewidth]{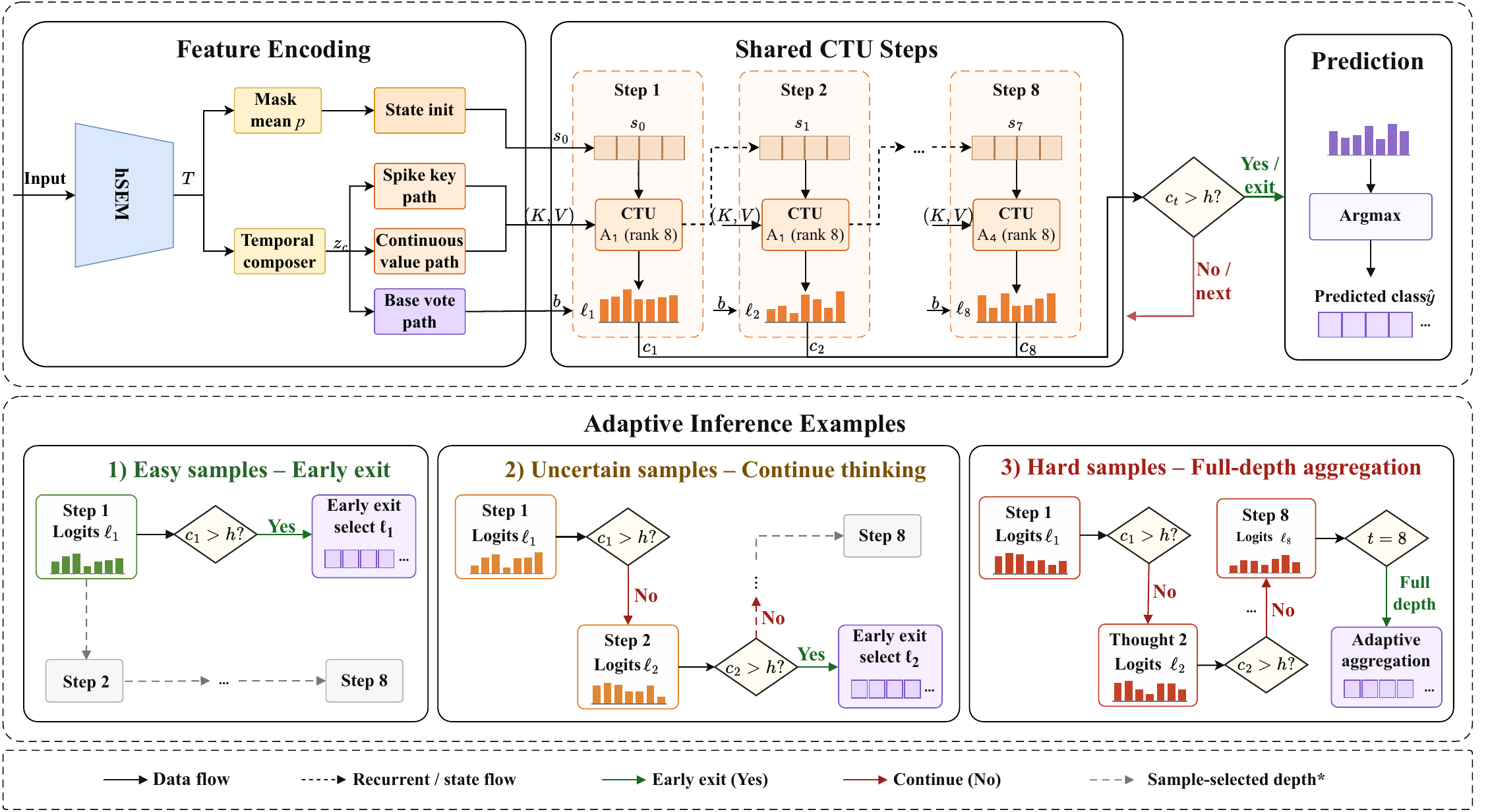}
  \caption{\method\ separates a fixed memory that preserves information from shared recurrent computation over a binary spike state. Each step emits evidence tied to the current prediction; inference either exits using current evidence or aggregates all steps. Grey paths denote logically skipped steps.}
  \label{fig:pipeline}
  \vspace{-15pt}
\end{figure}

Adaptive computation enables computational effort to scale with sample difficulty rather than a fixed worst-case budget. Adaptive Computation Time (ACT)~\citep{graves2016act}, PonderNet~\citep{banino2021pondernet}, and Continuous Thought Machines (CTM)~\citep{darlow2025ctm} make iteration count a computational dimension, and SNNs with early exit suppress redundant timesteps by temporal confidence~\citep{li2023seenn}. However, adaptive recurrence is not domain-neutral in SNNs. Binarizing membrane potentials into spikes limits the information carried by each neuron~\citep{bal2025pspikessm}, whereas a continuous path abandons the sparse, event-driven communication behind SNN efficiency~\citep{nunes2022spiking}. Recurrent spike state dynamics can also cause predictions to fluctuate across steps, such that confidence-based stopping may terminate prematurely on an incorrect prediction~\citep{li2023dynconf,chen2024knowing}, while relying solely on the final state can obscure both late corrections and late degradation. Thus, the key challenge is not simply to add recurrence to an SNN, but to organize continuous information, spike-domain computation, and stopping decisions so that each step remains useful and auditable.

Inspired by the internal computation in CTM, we propose \method, a compact spiking architecture guided by three design principles. \emph{Information preservation} retains continuous values in a fixed memory for repeated access. \emph{Recurrent computation over spikes} uses a shared Continuous Thinking Unit (CTU) to update a binary spike state while maintaining membrane dynamics across steps. \emph{Adaptive stopping from intermediate predictions} allows computation to end using predictions available at the current step. Together, these principles separate input encoding from recurrent computation and allow the number of recurrent steps to vary across samples. The resulting models combine competitive accuracy with few parameters and low estimated inference energy, addressing the memory and energy constraints of speech command recognition on edge devices.

Concretely, as illustrated in Fig.~\ref{fig:pipeline}, \method\ combines a hybrid spiking embedding module (hSEM), a temporal memory with keys obtained from spikes and continuous values, and the shared CTU with low-rank stage adapters. Each recurrent step consists of one memory read, one state update, and one intermediate prediction from the shared CTU. Aggregating over all steps learns a lightweight mixture over the resulting trajectory, whereas early exit returns the first sufficiently certain prediction based on the fixed base evidence without accessing future steps.
On the Spiking Heidelberg Digits, Spiking Speech Commands \citep{cramer2020heidelberg}, and Google Speech Commands V2 \citep{warden2018speechcommands} benchmarks, our method achieves competitive accuracy with 0.16--0.17\,M parameters. Our contributions are as follows:
\begin{itemize}
    \item We propose a compact spiking network for energy-efficient speech command recognition. It retains continuous values in a fixed memory and reuses a shared CTU to update a binary spike state, allowing additional recurrent computation with little parameter growth.
    \item We design an output mechanism that refines a fixed base prediction at each recurrent step. These intermediate predictions support aggregation at full depth and early exit based on current confidence.
    \item Experiments on SHD, SSC, and GSC show competitive accuracy with few parameters. Further analysis examines how predictions change across recurrent steps and shows that early exit improves accuracy over a baseline with the same average number of steps on SSC and GSC.
\end{itemize}

\section{Related Work}
\label{sec:related}

\paragraph{Spiking neural networks.}
SNNs transmit information through sparse, binary spike events rather than continuous values, and their event-driven dynamics promise low-power inference on neuromorphic hardware~\citep{nunes2022spiking}. Modern SNNs are typically trained directly by surrogate-gradient backpropagation through stateful neuron dynamics such as leaky integrate-and-fire (LIF) neurons~\citep{dampfhoffer2024backpropagation}, with learnable membrane constants adapting neuronal timescales~\citep{fang2021plif}. Recent SNNs have achieved competitive results in vision~\citep{zhou2023spikformer,yao2023spikedriven} and low-power neuromorphic deployment~\citep{bos2024micropower,wang2025background}. For instance, Spikformer V2 attains accuracy competitive with ANN Transformers on ImageNet-1K~\citep{zhou2024spikformerv2}. Nevertheless, the realized efficiency of such models depends on how information is represented, since binarizing continuous input values into spikes discards information, whereas a fully continuous path abandons the sparse communication behind SNN efficiency~\citep{hu2024largescale,sabbella2025promise}. This motivates our design, which maintains continuous input values in a fixed memory and recurrent computation over a binary spike state.

\paragraph{Spiking speech command recognition.}
Spiking SCR models process speech inputs using learned filterbanks~\citep{song2024spikingleaf} or convert them into events through temporal, pulse-density, and time-difference encoding~\citep{nilsson2023comparison,yarga2024neuromorphic,pequenozurro2025towards}. Downstream temporal structure is modeled through sparse recurrence~\citep{dampfhoffer2023leveraging}, temporal delay connections~\citep{malettira2025tskips,bouhadjar2026sparse}, structured-state and state-space neurons~\citep{bal2025pspikessm,ahmed2026spiking}. Concretely, SpikeSCR~\citep{wang2024efficient} obtains compact models through curriculum-based distillation, while SpikCommander~\citep{wang2026spikcommander} achieves strong SCR accuracy with a spiking embedding module and multi-view temporal attention. Efficiency has been pursued through early decisions, as in ED-sKWS~\citep{song2024edskws}, activity regularization~\citep{louis2024reducing}, compact delayed convolutions~\citep{li2026lightweight}, and asynchronous neuromorphic processors~\citep{zhang2024anpi}. Our focus is on adapting recurrent depth after sensory encoding, while retaining continuous memory values for repeated retrieval.

\paragraph{Conditional computation and adaptive inference.}
Conditional computation allocates compute to inputs that need it~\citep{bengio2013estimating,shazeer2017moe}. In ANNs, ACT~\citep{graves2016act}, PonderNet~\citep{banino2021pondernet}, Universal Transformers~\citep{dehghani2019universal}, and CTM~\citep{darlow2025ctm} establish recurrent depth as a controllable computation axis, but they do not address the placement of continuous information and binary state in an SNN. In SNNs, confidence-based policies shorten the timesteps of easy samples~\citep{li2023seenn,li2023dynconf,li2023dtsnn}, reinforcement learning guides the early exits~\citep{patne2026sparq}, and gating combines the outputs of spiking experts~\citep{yang2026spikingmoe,shim2016itdp}. Heuristic confidence thresholds can stop too early. SpikeCP instead uses conformal prediction to provide reliability guarantees at adaptive stopping times~\citep{chen2024knowing}. These approaches adapt simulation time, network depth, or expert selection. Our focus is on adapting recurrent computational depth after sensory encoding, using a shared unit to refine predictions without extending the input sequence.

\section{Method}
\label{sec:method}

As shown in Fig.~\ref{fig:pipeline}, \method\ first builds a fixed memory from the padded event sequence $X$, then repeatedly reads it with a shared recurrent unit. First, the hSEM encodes $X$ into hybrid tokens $T$, which are converted into the initial recurrent state $s_0$ and the fixed memory $C$. Then, $C$ yields spike keys $K$, continuous values $V$, and fixed base evidence $b$. One shared CTU projects the current state to a query and reads $K$ and $V$ repeatedly. Each cycle of this unit, comprising one memory read, one update of the recurrent state, and one emission of refined evidence, is a recurrent step. The number of steps $D$ defines recurrent depth. Every step updates a binary recurrent state $s_t$ and produces a prediction based on the fixed base evidence. Stopping is causal across steps, although the fixed memory encodes the complete input.
% We adopt LIF neurons with update $U[t]=H[t-1]+X[t]$, spike $S[t]=\Theta(U[t]-u_{\mathrm{th}})$, and reset $H[t]=\beta U[t](1-S[t])+V_{\mathrm{reset}}S[t]$, where $\Theta$ is the Heaviside step function and $\beta$ denotes a per-neuron decay.

\subsection{Fixed memory that preserves information}
\label{sec:memory}

Binarizing the input features into spikes discards their continuous information, and a continuous recurrent path loses efficiency from sparse spikes. The hSEM therefore maps the input to token width $d_T$ by combining two LIF spike paths with a gated continuous residual. Specifically, $X$ is embedded into a hidden continuous representation by pointwise and depthwise convolutions followed by batch normalization. The masked version of this representation gives the continuous residual $A$, while the same representation is converted into $S^{(1)}$ by a LIF, and $S^{(1)}$ then passes through a linear layer, batch normalization, and another LIF to yield $S^{(2)}$. The three branches are fused into the hybrid token:
\begin{equation}
  T = M\odot\big(S^{(1)}+S^{(2)}+\omega(a)A\big),
  \label{eq:token}
\end{equation}
where $\omega(\cdot)$ is the sigmoid function and $a$ is learnable, with the gate $\omega(a)$ initialized to $0.05$ to scale only the continuous residual. The masked mean of $T$ yields the pooled token $p=\frac{\sum_l T_l}{\sum_l M_l}$ that initializes the recurrent state, where $l$ indexes token positions and $M_l$ masks padded positions.

Then $T$ is mapped to state width $d_s$ via a linear layer $W$ followed by LayerNorm ($\mathrm{LN}$). Temporal composition then adds local and long-range context through depthwise branches with kernel sizes $\mathcal{K}=\{7,31\}$ and learnable gating scales $\varepsilon_k$. Hence, the fixed memory $C$ is given by:
\begin{equation}
  C = \mathrm{LN}(WT) + \sum_{k\in\mathcal{K}}\omega(\varepsilon_k)\,\mathrm{LN}_k(\mathrm{DWConv}_k(\mathrm{LN}(WT))),
  \label{eq:compose}
\end{equation}
The shared CTU reads this memory selectively. A token PLIF turns $C$ into keys obtained from spikes $K$, while the continuous $C$ supplies values $V$, so retrieval stays spike-based without binarizing the content. A head that averages token predictions on $C$ then produces the base evidence $b=\log\frac{\sum_l M_l\,\mathrm{softmax}(W_b C_l)}{\sum_l M_l}$, anchoring later predictions to a reference independent of the recurrent state.

\subsection{Shared recurrent computation over a binary spike state}
\label{sec:ctu}

\begin{figure}[t]
  \centering
  \includegraphics[width=\linewidth]{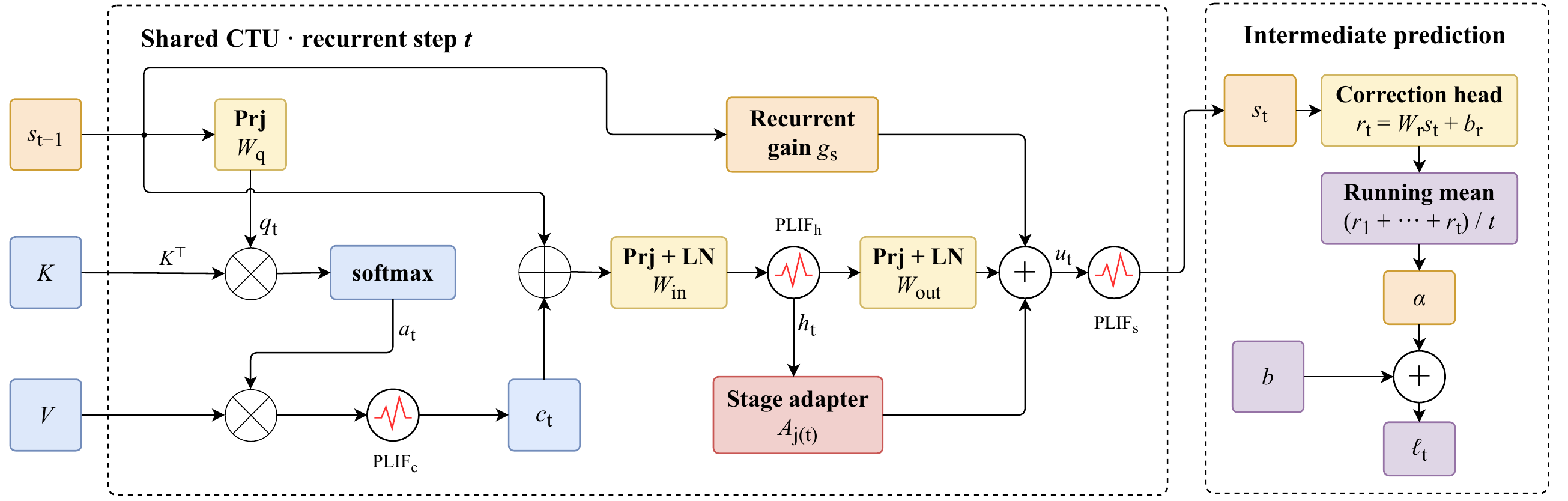}
  \caption{One recurrent CTU step. A query from the binary state $s_{t-1}$ reads fixed spike keys $K$ and continuous values $V$ with multi-head attention. The shared refiner updates $s_t$ with recurrent gain and a stage adapter. A shared classifier adds the running correction mean to base evidence $b$, producing $\ell_t$ and a causal confidence signal.}
  \label{fig:ctu}
  \vspace{-5pt}
\end{figure}
CTM uses neuron-level history models and synchronization representations~\citep{darlow2025ctm}. Our shared CTU instead reads the fixed memory using a binary state and updates it through PLIF dynamics (Fig.~\ref{fig:ctu}). The pooled token is mapped to state width $d_s$ and initializes $s_0\in\{0,1\}^{B\times d_s}$ via the shared state PLIF (Appendix~\ref{app:reproducibility}). At step $t$, $s_{t-1}$ is projected to a query $q_t = W_qs_{t-1}$ and reads the fixed $K$ and $V$ with multi-head attention, masking padded positions. The resulting context passes through a PLIF to obtain a spike-based $c_t$, which is concatenated with $s_{t-1}$ and fed into a shared refiner of width $2d_s$. The refiner first produces $h_t = \mathrm{PLIF}_{h}(\mathrm{LN}(W_{\mathrm{in}}[s_{t-1};c_t]))$ and then combines its output with a gated state connection and a stage adapter, giving
\begin{equation}
  u_t = \mathrm{LN}(W_{\mathrm{out}}h_t)+g_s s_{t-1}+A_{j(t)}(h_t),
  \label{eq:state}
\end{equation}
followed by $s_t = \mathrm{PLIF}_{s}(u_t)$. Here $g_s$ is softplus-parameterized to stay positive, and $W_{\mathrm{in}}, W_{\mathrm{out}}$ are learnable refiner projections. To give later stages limited dedicated capacity without unsharing the CTU, the rank-$r$ adapter $A_{j(t)}$ changes every two steps (one stage), while the reader, synapses, classifier, gains, and PLIF nodes remain shared. Increasing $D$ thus adds recurrent evaluations and only low-rank adapter capacity. The resulting state $s_t$ is passed to the shared classifier, which emits a correction to the fixed base evidence and supplies the confidence signal used for causal stopping, as shown in Fig.~\ref{fig:ctu}. Fig.~\ref{fig:reasoning_dynamics}(a) confirms that the binary activity pattern of $s_t$ changes across steps rather than replaying a fixed state.

\subsection{Intermediate predictions and adaptive output}
\label{sec:readout}

Persistent membrane dynamics can correct an early belief but can also damage it, so a last-state classifier hides both outcomes. Each step starts from the same base evidence and adds a running correction, giving $\ell_t=b+\alpha\,\frac{1}{t}\sum_{i=1}^{t}r_i$, where $r_t=W_rs_t+b_r$ and $\alpha$ is learnable and initialized small. The logits $\ell_t$ provide a causal confidence signal through normalized negative entropy, $q_t^{\mathrm{cert}}=1-H(\mathrm{softmax}(\ell_t))/\log N_c$, where $H$ is the Shannon entropy.

\begin{figure}[t]
  \centering
  \includegraphics[width=1.0\linewidth, trim=10 5 10 10]{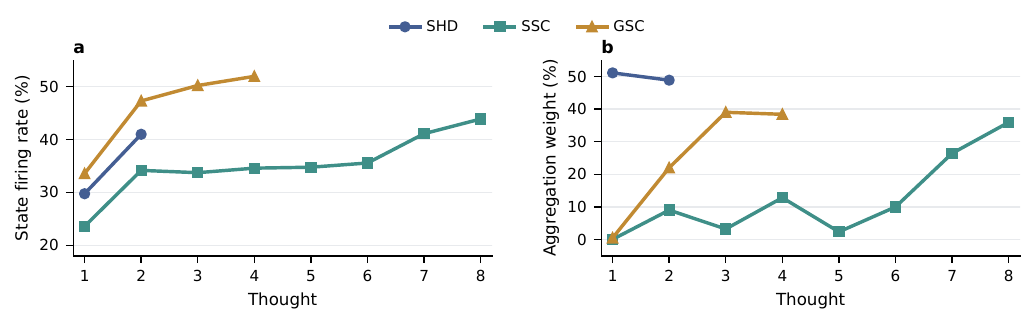}
  \caption{Diagnostics of the two learned mechanisms in the CTU. (a) Binary state activity varies across steps, indicating that the recurrent state is updated rather than replayed. (b) Aggregation assigns weights that vary across steps and tasks.}
  \label{fig:reasoning_dynamics}
  \vspace{-5pt}
\end{figure}

At full depth, scores $e_t=w^\top s_t+\rho_t+\gamma q_t^{\mathrm{cert}}$ combine learned state weights $w$, position biases $\rho_t$, and certainty weight $\gamma$, with $\pi_t=\mathrm{softmax}(e_1,\ldots,e_D)_t$. A learned $\beta$, initialized small, scales the correction to the last prediction:
\begin{equation}
  \ell_{\mathrm{full}}=\ell_D+\beta(m-\ell_D),
  \qquad m=\sum_{t=1}^{D}\pi_t\ell_t.
  \label{eq:full_output}
\end{equation}
Early exit returns the first $\ell_t$ satisfying $q_t^{\mathrm{cert}}>h$ (patience one), or $\ell_D$ otherwise. It always bypasses aggregation, including at depth $D$, and uses no future steps. Fig.~\ref{fig:reasoning_dynamics}(b) shows that aggregation weights vary across steps and tasks.

\subsection{Training objective}
\label{sec:objective}

Cross-Entropy (CE) on the output after all steps alone supervises only the final output, $\mathcal{L}_{\mathrm{full}}=\mathrm{CE}(\ell_{\mathrm{full}},y)$. To make every step useful, we add anytime supervision $\mathcal{L}_{\mathrm{any}}=\sum_{t=1}^{D}\omega_t\mathrm{CE}(\ell_t,y)$ whose weights assign greater total weight to later steps. Let $\varepsilon_{n,t}=\mathrm{CE}(\ell_{n,t},y_n)$, and let $\mathcal{C}_t$ and $\mathcal{W}_t$ be the samples correct and incorrect at step $t$. To preserve initially correct predictions, $\mathcal{L}_{\mathrm{pre}}$ penalizes increases in their CE relative to step 1. For later refinement, $\mathcal{L}_{\mathrm{ref}}$ minimizes CE on previously incorrect samples and penalizes CE increases on previously correct ones. Here $\mathrm{avg}$ denotes the mean over each indicated subset:
\begin{align}
\mathcal{L}_{\mathrm{pre}} &= \mathrm{avg}_{n\in\mathcal{C}_1,\,t>1}[\varepsilon_{n,t}-\varepsilon_{n,1}]_+,\\
\mathcal{L}_{\mathrm{ref}} &= \frac{1}{D-1}\sum_{t=2}^{D}\left(\mathrm{avg}_{n\in\mathcal{W}_{t-1}}\varepsilon_{n,t}+\mathrm{avg}_{n\in\mathcal{C}_{t-1}}[\varepsilon_{n,t}-\varepsilon_{n,t-1}]_+\right).
\end{align}
Both losses are zero for $D=1$, and empty subsets contribute zero. The total loss is $\mathcal{L}=\mathcal{L}_{\mathrm{full}}+\lambda_{\mathrm{any}}\mathcal{L}_{\mathrm{any}}+\lambda_{\mathrm{pre}}\mathcal{L}_{\mathrm{pre}}+\lambda_{\mathrm{ref}}\mathcal{L}_{\mathrm{ref}}$, with the coefficients and task-specific switches given in Appendix~\ref{app:reproducibility}.

\section{Experimental Setups}
\label{sec:setup}

\paragraph{Datasets, protocols, and optimization.}
We evaluate \method\ on the 20-class Spiking Heidelberg Digits (SHD), the 35-class Spiking Speech Commands (SSC), and the 35-class Google Speech Commands V2 (GSC)~\citep{cramer2020heidelberg,warden2018speechcommands}. SHD provides 8,156 training and 2,264 test utterances without an official validation split, so we follow the standard SHD protocol. SSC uses 75,466/9,981/20,382 samples, and GSC uses the official 84,843/9,981/11,005 train/validation/test splits. For SSC and GSC, we select the best performing model on validation sets and evaluate it on test sets; SHD follows its standard train/test protocol. For fair comparison, we set the temporal resolution $\T=100$. All \method\ models use token width $d_T=128$, state width $d_s=96$, hidden ratio 2, $H=8$ reader heads, rank $r=8$ stage adapters, hSEM dropout 0.15, and AdamW with a cosine schedule. More implementation details and hyperparameters can be found in Appendix~\ref{app:reproducibility}.

\paragraph{Baselines.}
We compare with DCLS-Delays~\citep{hammouamri2024learning}, d-cAdLIF~\citep{deckers2024colearning}, SpikeSCR~\citep{wang2024efficient}, Pfa-SNN~\citep{sun2025pfa}, SiLIF~\citep{fabre2025silif}, DelRec~\citep{queant2025delrec}, EventProp~\citep{meszaros2025efficient}, Spiking LMUFormer~\citep{liu2024lmuformer}, SpikCommander~\citep{wang2026spikcommander}, and the ANN baseline KW-MLP~\citep{morshed2021attentionfree}. DCLS-Delays and d-cAdLIF results come directly from their original papers. Baseline energy estimates follow SpikCommander's reported comparisons. We retain published configurations rather than retraining baselines with identical preprocessing. Source details are given in Appendix~\ref{app:baseline_sources}.

\paragraph{Metrics.}
We report top-1 accuracy and trainable parameters for predictive performance and compactness. For computational cost, we report sparse synaptic operations (SOPs) for spiking layers and multiply--accumulate operations (MACs) for continuous paths, together with a 45-nm proxy energy $E=0.9\,\mathrm{pJ}\times\mathrm{AC}+4.6\,\mathrm{pJ}\times\mathrm{MAC}$, following prior operation-based estimates. 
% For adaptive inference we additionally report the number of executed steps. These are analytical proxies that exclude memory movement, control, and other device-level overheads.

\section{Results}
\label{sec:results}

\subsection{Accuracy, model size, and operation estimates}

\begin{table}[t]
\vspace{-15pt}
\centering
\small
\setlength{\tabcolsep}{4.2pt}
\renewcommand{\arraystretch}{0.82}
\caption{Accuracy, trainable parameters, and theoretical 45-nm proxy energy. \textsuperscript{A} denotes ANN. Bold and underlined values rank first and second within each dataset. ``--'' denotes unavailable results. Baseline sources are given in Appendix~\ref{app:baseline_sources}, and multi-seed results for \method\ in Appendix~\ref{app:multiseed}.}
\label{tab:main_comparison}
\begin{tabular}{llrrr}
\toprule
Data & Model & Params (M) & Acc. (\%) & Energy (mJ) \\
\midrule
\multirow{7}{*}{SHD}
% Primary sources: Hammouamri et al., Table 2 (2L-1KC); Deckers et al., Table 3.
 & DCLS-Delays (2L) \citep{hammouamri2024learning} & 0.20 & 95.07 & -- \\
 & d-cAdLIF \citep{deckers2024colearning} & \textbf{0.08} & 94.85 & -- \\
 & SpikeSCR (1L) \citep{wang2024efficient} & 0.26 & 95.60 & \underline{0.0051} \\
 & Pfa-SNN \citep{sun2025pfa} & 0.20 & 96.26 & -- \\
 & EventProp+Delays \citep{meszaros2025efficient} & -- & 93.20 & -- \\
 & SpikCommander (1L) \citep{wang2026spikcommander} & 0.19 & \underline{96.41} & \textbf{0.0039} \\
\rowcolor{blue!7}
 & \method\ ($D=2$) & \underline{0.16} & \textbf{96.60} & 0.0140 \\
\midrule
\multirow{9}{*}{SSC}
% DCLS-Delays uses the 3L-2KC configuration from the original Table 2.
 & DCLS-Delays (3L) \citep{hammouamri2024learning} & 2.50 & 80.69 & -- \\
 & d-cAdLIF \citep{deckers2024colearning} & 0.70 & 80.23 & -- \\
 & SpikeSCR (2L) \citep{wang2024efficient} & 3.30 & \underline{82.79} & 0.0314 \\
 & DelRec \citep{queant2025delrec} & 0.37 & 82.60 & -- \\
 & SiLIF \citep{fabre2025silif} & \underline{0.35} & 82.03 & -- \\
 & Pfa-SNN \citep{sun2025pfa} & 0.71 & 80.18 & -- \\
 & EventProp+Delays \citep{meszaros2025efficient} & -- & 76.10 & -- \\
 & SpikCommander (2L) \citep{wang2026spikcommander} & 2.13 & \textbf{83.49} & \underline{0.0180} \\
\rowcolor{blue!7}
 & \method\ ($D=8$) & \textbf{0.17} & 81.74 & \textbf{0.0158} \\
\midrule
\multirow{8}{*}{GSC}
 & KW-MLP\textsuperscript{A} \citep{morshed2021attentionfree} & \underline{0.42} & \textbf{97.56} & 0.2070 \\
% DCLS-Delays uses the 3L-2KC configuration from the original Table 2.
 & DCLS-Delays (3L) \citep{hammouamri2024learning} & 2.50 & 95.35 & -- \\
 & d-cAdLIF \citep{deckers2024colearning} & 0.61 & 95.69 & -- \\
 & SpikeSCR (1L) \citep{wang2024efficient} & 1.63 & 95.56 & 0.0670 \\
 & Spiking LMUFormer \citep{liu2024lmuformer} & 1.69 & 96.12 & 0.0590 \\
 & SiLIF+Delays \citep{fabre2025silif} & 0.61 & 95.88 & -- \\
 & SpikCommander (2L) \citep{wang2026spikcommander} & 2.13 & \underline{96.92} & \underline{0.0420} \\
\rowcolor{blue!7}
 & \method\ ($D=4$) & \textbf{0.17} & 95.96 & \textbf{0.0231} \\
\bottomrule
\end{tabular}
\end{table}

\method\ has the smallest reported parameter count on SSC and GSC and the highest reported accuracy on SHD among the compared methods (Table~\ref{tab:main_comparison}). On SHD, d-cAdLIF is smaller, while our 96.60\% accuracy exceeds SpikCommander by 0.19 percentage points under a seed-matched protocol. On SSC, it trails the best-performing model, SpikCommander, by 1.75 points while using approximately one twelfth of its parameters. On GSC, it is 0.96 points below SpikCommander, the most accurate spiking baseline, while using only 55\% of its reported energy. Among methods with energy estimates, \method\ has the lowest reported values on SSC and GSC. These comparisons support a compact design with competitive accuracy. Energy rankings differ on SHD, where some baselines count all synaptic operations as accumulations, while we count continuous paths as MACs. The accounting conventions and operation breakdown are given in Appendix~\ref{app:energy}.

\subsection{Ablation studies}
\begin{table}[t]
\vspace{-15pt}
\centering
\caption{Cross-task component ablation. Each row removes or replaces one component of the full model on the first row. Bold and underline mark the best and second-best accuracy per task. $^{\dagger}$The full SHD model trains without $L_{\mathrm{ref}}$, so this row enables it on SHD.}
\label{tab:cross_task_ablation}
\small
\setlength{\tabcolsep}{4.5pt}
\renewcommand{\arraystretch}{0.95}
\begin{tabular}{lcccccc}
\toprule
\multirow{2}{*}{Configuration} & \multicolumn{2}{c}{SHD ($D{=}2$)} & \multicolumn{2}{c}{SSC ($D{=}8$)} & \multicolumn{2}{c}{GSC ($D{=}4$)} \\
\cmidrule(lr){2-3}\cmidrule(lr){4-5}\cmidrule(lr){6-7}
 & Acc. (\%) & Params (K) & Acc. (\%) & Params (K) & Acc. (\%) & Params (K) \\
\midrule
Full & \textbf{96.60} & 164.23 & \textbf{81.74} & 174.04 & \textbf{95.96} & 169.43 \\
\midrule
w/o temporal composition & 94.30 & 160.19 & 78.99 & 170.01 & 94.38 & 165.39 \\
w/ continuous K/V memory & 95.14 & 164.23 & 81.28 & 174.04 & 95.85 & 169.43 \\
w/ spike K/V memory & 95.49 & 164.23 & 81.12 & 174.04 & 95.75 & 169.43 \\
w/o stage adapters & 94.83 & 161.92 & 81.24 & 164.82 & 95.85 & 164.82 \\
w/o adaptive aggregation & 94.43 & 164.13 & 81.21 & 173.94 & 95.85 & 169.33 \\
w/o $L_{\mathrm{any}}$ & \underline{95.85} & 164.23 & 81.12 & 174.04 & \underline{95.90} & 169.43 \\
w/o $L_{\mathrm{pre}}$ & 95.63 & 164.23 & \underline{81.49} & 174.04 & 95.57 & 169.43 \\
w/o $L_{\mathrm{ref}}$ & 96.16$^{\dagger}$ & 164.23 & 81.63 & 174.04 & 95.71 & 169.43 \\
\bottomrule
\end{tabular}
\vspace{-5pt}
\end{table}

Temporal composition makes the largest consistent contribution across tasks. Removing it reduces accuracy by 1.58--2.75 percentage points while saving only about 4 K parameters (Table~\ref{tab:cross_task_ablation}). Replacing the hybrid memory with either continuous keys and values or spiking keys and values also lowers accuracy at unchanged model size. These results support combining temporal context with continuous memory values and keys obtained from spikes. Stage adapters and adaptive aggregation provide further gains with limited parameter overhead. The adapters allow some specialization across recurrent stages while retaining the shared CTU. Aggregation improves on using only the last prediction, gaining 2.17 percentage points on SHD with about 0.1 K additional parameters. Thus, even a short recurrent sequence can benefit from combining intermediate predictions. Anytime supervision and preservation of the first prediction improve accuracy across all three tasks without adding parameters. Refinement loss has a less uniform effect. It helps on SSC and GSC but lowers accuracy on SHD, supporting the task-specific choice used in the full models.

\subsection{Recurrent updates can correct or degrade predictions}
\label{sec:results:reasoning}
\begin{figure}[htbp]
  \centering
  \includegraphics[width=1.0\linewidth, trim=0 10 0 10, clip]{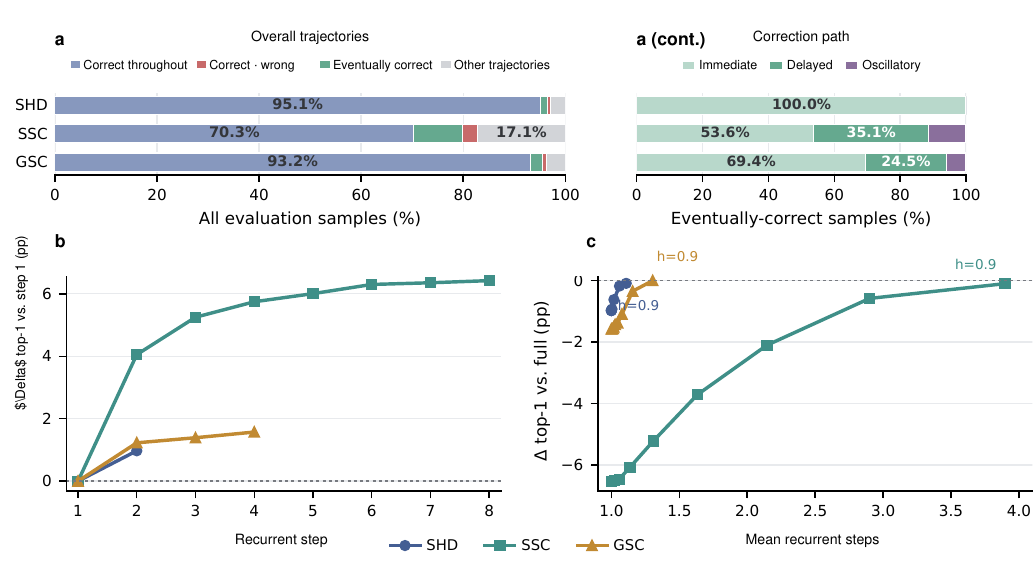}
  \caption{Prediction changes during recurrent computation. (a) Trajectory composition and correction types. Detailed counts are given in Appendix~\ref{app:trajectories}. (b) Top-1 change across steps of the same model. (c) Top-1 gap to full inference under $h{=}0.9$ early exit.}
  \label{fig:reasoning}
  \vspace{-10pt}
\end{figure}

\begin{figure}[htbp]
  \centering
  \includegraphics[width=\linewidth]{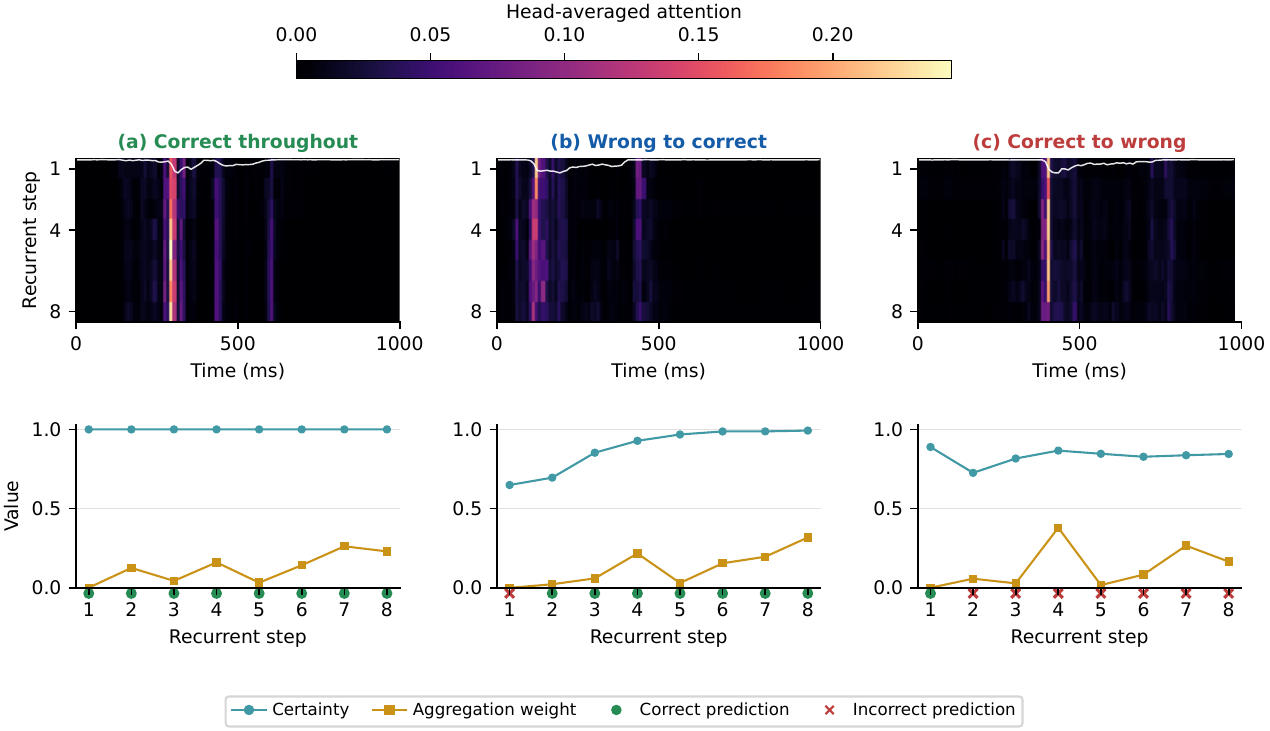}
  \caption{Sample-level attention across recurrent steps on SSC. Columns show predictions that (a) remain correct, (b) change from wrong to correct, and (c) change from correct to wrong. Top: attention maps. Bottom: certainty, aggregation weight, and correctness.}
  \label{fig:attention_examples}
  \vspace{-5pt}
\end{figure}

Average accuracy improves across recurrent steps, but individual predictions can change in either direction (Fig.~\ref{fig:reasoning}). On SSC, 9.48\% of samples change from an incorrect initial prediction to a correct final prediction, while 3.06\% change in the opposite direction. Some corrections occur only after several steps or after repeated changes in correctness. These trajectories show why the last prediction alone provides an incomplete view of recurrent computation. Panel (b) follows successive outputs of one trained model. It should be distinguished from the depth sweep, where a separate model is trained at each depth. Detailed trajectory counts are provided in Appendix~\ref{app:trajectories}. The state and aggregation diagnostics in Fig.~\ref{fig:reasoning_dynamics} show that recurrent steps update the binary state and receive different aggregation weights. Figure~\ref{fig:attention_examples} illustrates attention across recurrent steps for three SSC examples whose predictions remain correct, change from wrong to correct, or change from correct to wrong. Attention also moves in both corrected and degraded predictions, so we examine temporal occlusion alongside attention weights~\citep{jain2019attention,wiegreffe2019attention,serrano2019attention}. On SSC, masking the top 10\% of attended bins at the final step reduces accuracy by 17.65 percentage points, compared with 4.55 points for random masking of the same proportion. Together, these results indicate that the model uses the attended input regions, while changes in attention alone do not establish that a prediction has improved (Appendix~\ref{app:attention}).

\subsection{Adaptive computation with selected thresholds}

\begin{wraptable}{r}{0.55\columnwidth}
\vspace{-22pt}
\centering
\caption{Adaptive computation with thresholds selected on validation data: accuracy, mean evaluated steps, and gain over a fixed-depth result at the same average compute.}
\label{tab:early_exit}
\scriptsize
\setlength{\tabcolsep}{3.5pt}
\renewcommand{\arraystretch}{0.90}
\begin{tabular}{lcccccc}
\toprule
Data & $D$ & $h$ & Acc. (\%) & Mean steps & Reduction & Gain (pp) \\
\midrule
SHD & 2 & 0.01 & 95.63 & 1.00 & 50.00\% & +0.00 \\
SSC & 8 & 0.84 & 81.42 & 3.28 & 59.04\% & +0.81 \\
GSC & 4 & 0.91 & 95.99 & 1.32 & 66.97\% & +1.21 \\
\bottomrule
\end{tabular}
\vspace{-6pt}
\end{wraptable}

Early exit reduces recurrent computation by 59\% on SSC and 67\% on GSC (Table~\ref{tab:early_exit}). At the same average number of steps, it improves accuracy over the interpolated baseline by 0.81 and 1.21 percentage points, respectively. SHD shows no gain over this baseline at its shallow selected depth. The comparison suggests that intermediate confidence can allocate recurrent computation more effectively than uniform truncation on SSC and GSC. Thresholds are selected on calibration data with an allowed accuracy loss of at most 0.1 percentage points. The baseline interpolates between adjacent truncation depths of the same frozen model, and exited predictions use no future steps. Initial prediction uncertainty is also informative about exit depth, although a coarse allocation based only on this uncertainty performs less well than confidence stopping (Appendix~\ref{app:difficulty}). The reported savings concern recurrent evaluations. Encoding and memory construction remain fixed and dominate estimated energy, so the reduction in total inference energy is smaller (Appendix~\ref{app:energy}).

\subsection{Depth and resolution depend on the task}
\begin{wraptable}{r}{0.49\linewidth}
\vspace{-20pt}
\centering
\caption{Top-1 accuracy (\%) at different sensory resolutions. $\Delta=\mathrm{Acc}_{250}-\mathrm{Acc}_{100}$, in percentage points.}
\label{tab:time_steps_compact}
\small
\setlength{\tabcolsep}{2.7pt}
\renewcommand{\arraystretch}{0.82}
\begin{tabular}{@{}lcccc@{}}
\toprule
Data & $\T=50$ & $\T=100$ & $\T=250$ & $\Delta$ \\
\midrule
SHD & 95.01 & \textbf{96.60} & 94.66 & $-1.94$ \\
SSC & 79.78 & 81.74 & \textbf{82.53} & $+0.79$ \\
GSC & 95.25 & \textbf{95.96} & 95.44 & $-0.52$ \\
\bottomrule
\end{tabular}
\vspace{-8pt}
\end{wraptable}

\begin{figure}[t]
  \centering
  \includegraphics[width=0.94\linewidth, trim=5 10 5 10, clip]{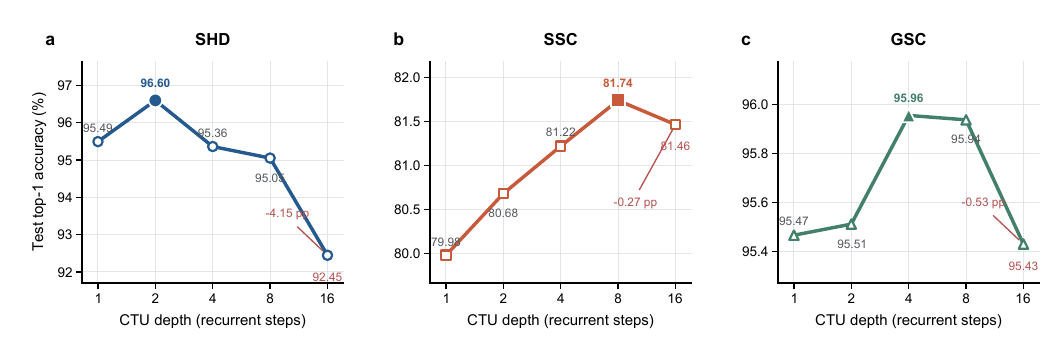}
  \caption{Top-1 accuracy versus recurrent CTU depth on SHD, SSC, and GSC. Each point is trained and evaluated at the indicated number of steps. The different optima show that additional recurrent computation does not always improve accuracy.}
  \label{fig:depth}
  % \vspace{-5pt}
\end{figure}

The best recurrent depths are $D=2,8,4$ for SHD, SSC, and GSC, respectively (Fig.~\ref{fig:depth}). Further increases reduce accuracy, most clearly on SHD, where $D=16$ falls 4.15 percentage points below the optimum. Because the CTU is shared, increasing depth adds repeated evaluations with only modest parameter growth. More recurrent computation is therefore useful only up to a task-dependent depth, and depth should be selected rather than increased by default. Sensory resolution has a different role. Table~\ref{tab:time_steps_compact} shows that increasing $\T$ from 50 to 100 improves all three tasks, whereas increasing it to 250 helps only SSC. While $\T$ controls the granularity and cost of input encoding, $D$ controls the number of recurrent updates after encoding. We use $\T=100$ for the main accuracy and energy comparisons and report the full sweep in Appendix~\ref{app:time_steps}. Each configuration uses one controlled run, so these sweeps describe sensitivity rather than statistical uncertainty.

\section{Limitations}
\label{sec:limitations}

Our study has three methodological limitations. First, the CTU reads a fixed, composed memory using softmax attention over keys obtained from spikes and continuous values. Future work could therefore investigate writable sparse memory and matrix-free reading, since adaptive depth does not remove the fixed cost of constructing and retaining the composed memory. Second, the halting rule is calibrated after training rather than learned under an explicit compute budget, and the hyperparameters of the recurrent loop are fixed rather than systematically swept. Post-hoc calibration may not transfer across sensory resolutions or deployment constraints. Third, our evidence is limited to supervised closed-set speech classification. Evaluation on streaming settings, open-vocabulary tasks, and other event modalities is needed to clarify the limits of transferability.

\section{Conclusion}
\label{sec:conclusion}

\method\ presents a compact way to organize information and recurrent computation in an SNN. Continuous values remain available in a fixed memory, while binary states support repeated updates and intermediate predictions support adaptive stopping. Across SHD, SSC, and GSC, the resulting models use 0.16--0.17 M parameters and achieve a favorable balance between accuracy and model size. Our analyses of prediction trajectories make the recurrent process visible, showing that later steps can correct an initial prediction, reinforce it, or occasionally degrade it. Calibrated stopping further shows that recurrent depth can be allocated to individual samples, improving the accuracy--computation trade-off on SSC and GSC. Sensitivity studies distinguish the roles of sensory resolution and recurrent depth, with $\T$ controlling the granularity of encoded evidence and $D$ controlling computation after encoding. Together, these results establish recurrent depth as a useful axis for adaptive spike-domain sequence computation and point toward policy-consistent training, writable memory, and hardware evaluation.

\bigskip
\bibliography{references}
\bibliographystyle{iclr2027_conference}

\clearpage
\section*{AI Use Disclosure}

We used generative AI tools to improve the readability and language of selected parts of the manuscript and to assist with the editing of figure labels and captions. We reviewed all AI-assisted text and visual edits manually and verified them against the implementation, experiments, and cited literature. The authors take full responsibility for the final content of this paper, including all claims, results, and artifacts.

\bigskip
\appendix
\section*{Supplementary Material}
\raggedbottom

\section{Reproducibility and Robustness}
\label{app:reproducibility}

\subsection{Task configurations and input construction}

Table~\ref{tab:training_config} records the settings for each task used for the reported models. Architecture width is fixed across tasks. Only recurrent depth, the number of adapter stages, optimizer settings for each task, and the refinement-loss switch differ. The seed used in the main paper, 312, is also one of the five robustness seeds.

\begin{table}[H]
\centering
\caption{Frozen task configurations. LR and WD denote the initial learning rate and weight decay, respectively.}
\label{tab:training_config}
\small
\begin{tabular}{lcccccccc}
\toprule
Data & Classes & $D$ & Adapters & Epochs & LR & WD & Batch & $\lambda_{\mathrm{ref}}$ \\
\midrule
SHD & 20 & 2 & 1 & 500 & 0.010 & 0.010 & 256 & 0.00 \\
SSC & 35 & 8 & 4 & 300 & 0.005 & 0.020 & 256 & 0.05 \\
GSC & 35 & 4 & 2 & 300 & 0.002 & 0.005 & 256 & 0.05 \\
\bottomrule
\end{tabular}
\end{table}

The pooled token passes through a linear projection from $d_T$ to $d_s$ and LayerNorm, then a second linear map and LayerNorm at width $d_s$. A learned seed normalized across state dimensions is added before the shared state PLIF produces $s_0$.

All tasks use token/state/hidden dimensions $128/96/192$, eight reader heads, rank-8 adapters with two steps per stage, adaptive aggregation initialized at 0.05, hSEM dropout 0.15, AdamW, and cosine annealing with period 40. Dropout is applied after each hSEM spike layer, and dropout in the CTU step loop is zero. Teacher, logit distillation, token-relation distillation, distillation from the final step, the auxiliary head that averages token predictions, and flip-loss branches are disabled. All LIF and PLIF states are reset after every batch. Thus, Table~\ref{tab:training_config} shows that model width and the recurrent mechanism are fixed across tasks. The main task-specific architectural variable is recurrent depth, with the number of adapter stages adjusted accordingly. The refinement switch is the only loss difference because the controlled SHD study finds it detrimental at $D=2$.

SHD and SSC use 10-ms event integration at the reported $\T=100$ setting. Cochlear indices are grouped in contiguous sets of five, reducing 700 sensors to 140 feature channels. Padding is zero and excluded by mask from pooling, attention, and token voting. Independent temporal and feature masks are sampled with probability 0.5. SHD masks at most 20\% of the time steps and 20 feature channels, while SSC uses 10\% of the time steps and 10 feature channels. GSC truncates or pads waveforms before the released 8-kHz log-Mel pipeline, uses $n_{\mathrm{fft}}=256$, a hop length of 80, and 140 Mel bins, then applies one frequency mask and one time mask. We preserve the released $f_{\max}=14{,}000$ setting despite the 8-kHz target sample rate to avoid silently changing the benchmark feature pipeline.

\subsection{Training losses and policy calibration}

For reproducibility, let $\varepsilon_{n,t}=\mathrm{CE}(\ell_{n,t},y_n)$ denote per-sample cross-entropy. Late anytime weights allocate total mass 0.1 uniformly to the first $d=\lfloor D/2\rfloor$ steps and mass 0.9 to the remaining steps in proportion to $v_i=\min(i,3)$. Thus, for $D>1$,
\begin{equation}
  \omega_t=\begin{cases}
    0.1/d, & t\le d,\\
    0.9v_{t-d}/\sum_{i=1}^{D-d}v_i, & t>d,
  \end{cases}
  \qquad \sum_t\omega_t=1,
  \label{eq:late_weights}
\end{equation}
while $\omega_1=1$ when $D=1$. The outer anytime coefficient is $\lambda_{\mathrm{any}}=0.1$.

Preservation of the first prediction operates only on samples predicted correctly at the first step. With the margin set to zero, it averages $[\varepsilon_{n,t}-\varepsilon_{n,1}]_+$ over those samples and $t>1$; its coefficient is $\lambda_{\mathrm{pre}}=0.05$. Belief refinement partitions samples at each $t>1$ by the correctness of step $t-1$. It averages $\varepsilon_{n,t}$ over samples that were previously incorrect and $[\varepsilon_{n,t}-\varepsilon_{n,t-1}]_+$ over samples that were previously correct. It then sums these two subset means and averages the result across later steps. Its preservation multiplier is one and $\lambda_{\mathrm{ref}}=0.05$ on SSC/GSC; SHD sets $\lambda_{\mathrm{ref}}=0$. An empty subset contributes zero, matching the released implementation.

SSC and GSC select thresholds using their full official validation splits. Because SHD has no official validation split, its policy for early exit instead uses a deterministic 20\% subset of the training data, stratified by class, for calibration (seed 20260730). Thresholds are selected under at most 0.1 percentage-point loss on calibration relative to the adaptive output after all steps, without observing evaluation labels. Patience is one and the condition is strict, $q_t^{\mathrm{cert}}>h$.

\subsection{Multi-seed robustness}
\label{app:multiseed}

\begin{table}[H]
\centering
\caption{Five independent runs. Values are top-1 accuracy (\%) under the official evaluation protocol; boldface and underlining indicate the best and second-best results, respectively, for each task. Tied second-best values are underlined.}
\label{tab:multiseed_full}
\small
\setlength{\tabcolsep}{5.5pt}
\begin{tabular}{lccccccc}
\toprule
Data & Params (M) & Seed 312 & Seed 326 & Seed 12306 & Seed 11408 & Seed 42 & Mean $\pm$ s.d. \\
\midrule
SHD & 0.16 & \textbf{96.60} & \underline{95.80} & 95.27 & 95.10 & 94.79 & $95.51\pm0.71$ \\
SSC & 0.17 & \textbf{81.74} & \underline{81.42} & 81.38 & 81.25 & 81.18 & $81.39\pm0.22$ \\
GSC & 0.17 & \textbf{95.96} & \underline{95.80} & 95.76 & 95.60 & \underline{95.80} & $95.78\pm0.13$ \\
\bottomrule
\end{tabular}
\end{table}

The five-run means are $95.51\pm0.71$\%, $81.39\pm0.22$\%, and $95.78\pm0.13$\% for SHD, SSC, and GSC, where the uncertainty is the sample standard deviation across seeds. Cross-seed ranges are 1.81, 0.56, and 0.35 points, respectively. Seed 312 exceeds the corresponding mean by 1.09 points on SHD, compared with 0.35 on SSC and 0.18 on GSC. Thus, the headline SHD result is more seed-sensitive, whereas the SSC/GSC models form tighter clusters around their means. SSC/GSC entries are independent-test evaluations of checkpoints selected on validation; SHD entries follow the official train/test reporting protocol used by prior work. These five runs quantify robustness; the single-seed depth, timestep, and ablation sweeps remain sensitivity evidence, not significance tests.

\subsection{Baseline sources and reporting conventions}
\label{app:baseline_sources}

DCLS-Delays uses 2L-1KC on SHD and 3L-2KC on SSC/GSC, the most accurate configurations in Table~2 of \citet{hammouamri2024learning}. The d-cAdLIF results come from Table~3 of \citet{deckers2024colearning}. For both methods, Table~\ref{tab:main_comparison} lists the published mean accuracies without uncertainty intervals, not the best individual runs. Parameter counts are rounded to two decimal places. Both use 10-ms spiking input bins, but GSC preprocessing differs (140 versus 40 Mel bins). Neither supplies comparable energy estimates for these configurations. Energy values for the remaining baselines follow Tables~2--3 of \citet{wang2026spikcommander}.

\section{Sensitivity and Adaptive Computation}

\subsection{Sensitivity to recurrent depth}

\begin{table}[H]
\centering
\caption{Recurrent-depth sensitivity (top-1 accuracy, \%). Boldface and underlining indicate the best and second-best results, respectively, for each task.}
\label{tab:depth_full}
\normalsize
\setlength{\tabcolsep}{8.5pt}
\renewcommand{\arraystretch}{1.16}
\begin{tabular}{@{}lccccc@{}}
\toprule
Data & $D=1$ & $D=2$ & $D=4$ & $D=8$ & $D=16$ \\
\midrule
SHD & \underline{95.49} & \textbf{96.60} & 95.36 & 95.05 & 92.45 \\
SSC & 79.98 & 80.68 & 81.22 & \textbf{81.74} & \underline{81.46} \\
GSC & 95.47 & 95.51 & \textbf{95.96} & \underline{95.94} & 95.43 \\
\bottomrule
\end{tabular}
\end{table}

The full numerical results in Table~\ref{tab:depth_full} correspond to Fig.~\ref{fig:depth}. Relative to $D=1$, the task-specific optima improve SHD, SSC, and GSC by 1.11, 1.76, and 0.49 points. Unrolling to $D=16$ instead loses 4.15, 0.28, and 0.53 points relative to those optima. The pronounced SHD decline, together with the shallower optima on SHD/GSC, shows that depth is not a uniform capacity multiplier. Each configuration is trained independently: a $D=1$ model is not obtained by truncating a $D=8$ model. Table~\ref{tab:depth_full} therefore measures the effects of training at each depth and recurrent dynamics, rather than optional post-training unrolling of an invariant predictor.

\begin{samepage}
\subsection{Sensitivity to sensory time steps}
\label{app:time_steps}

\noindent
\begin{minipage}[t]{0.485\linewidth}
\vspace{0pt}
SHD and GSC exhibit a saturating resolution regime, whereas SSC continues to benefit from finer temporal sampling. SHD rises from 87.94\% at $\T=10$ to 96.60\% at $\T=100$, then falls by 2.17 points at $\T=200$. GSC similarly improves from 85.31\% to 95.96\% at $\T=100$ before declining to 95.44\% at $\T=250$. In contrast, SSC rises from 66.75\% to 82.53\% across the sweep; its $\T=250$ optimum is 0.79 points above the common-resolution $\T=100$ result. Fig.~\ref{fig:time_steps} therefore supports task-specific sensory resolution rather than a monotonic cross-task rule.
SHD and SSC use integration windows of 100, 50, 25, 20, 10, 5, and 4 ms, corresponding to 10, 20, 40, 50, 100, 200, and 250 nominal steps. GSC instead fixes the spectrogram window at 256 samples and varies hop length. Thus, $\T$ changes sensory evidence granularity before recurrent CTU inference rather than changing the number of steps.
\end{minipage}\hfill
\begin{minipage}[t]{0.45\linewidth}
\vspace{0pt}
\centering
\includegraphics[width=\linewidth]{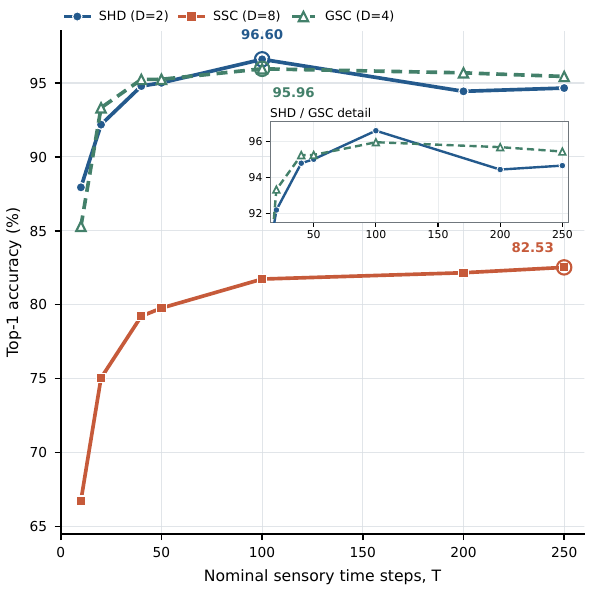}
\captionof{figure}{Sensory time-step sensitivity. Open markers show task-specific optima.}
\label{fig:time_steps}
\end{minipage}
\end{samepage}

Beyond $\T=100$, SHD and GSC decline while SSC continues to improve through $\T=250$. Temporal resolution is therefore task dependent rather than a universally monotonic compute dimension. Main comparisons nevertheless retain $\T=100$ so that input resolution and operation accounting remain fixed across tasks.

\begin{table}[H]
\centering
\caption{Sensory time-step sensitivity (top-1 accuracy, \%). Boldface and underlining indicate the best and second-best results, respectively, for each task.}
\label{tab:time_steps_full}
\small
\renewcommand{\arraystretch}{0.92}
\begin{tabular*}{\linewidth}{@{\extracolsep{\fill}}lccccccc@{}}
\toprule
Data & $\T=10$ & $\T=20$ & $\T=40$ & $\T=50$ & $\T=100$ & $\T=200$ & $\T=250$ \\
\midrule
SHD & 87.94 & 92.18 & 94.79 & \underline{95.01} & \textbf{96.60} & 94.43 & 94.66 \\
SSC & 66.75 & 75.05 & 79.23 & 79.78 & 81.74 & \underline{82.17} & \textbf{82.53} \\
GSC & 85.31 & 93.32 & 95.24 & 95.25 & \textbf{95.96} & \underline{95.68} & 95.44 \\
\bottomrule
\end{tabular*}
\end{table}

Table~\ref{tab:time_steps_full} provides the exact checkpoints underlying the compact plot. The different behavior at higher resolution is not compared vertically across datasets because class spaces, recurrent depths, and evaluation protocols differ. Every entry is one controlled run, so the sweep characterizes sensitivity rather than uncertainty.

\subsection{Adaptive computation and difficulty}
\label{app:difficulty}

\begin{table}[H]
\centering
\caption{Difficulty-aware allocation. Tertiles of normalized Step 1 entropy define the easy, medium, and hard groups; depths that do not decrease across these groups are selected on calibration data subject to the primary budget for early exit.}
\label{tab:difficulty}
\small
\begin{tabular}{lccc}
\toprule
Data & Depths (easy/medium/hard) & Accuracy (\%) & Mean steps \\
\midrule
SHD & $1/1/1$ & 95.63 & 1.00 \\
SSC & $1/3/5$ & 81.12 & 3.06 \\
GSC & $1/1/2$ & 95.61 & 1.35 \\
\bottomrule
\end{tabular}
\end{table}

Difficulty-aware allocation partitions calibration samples into tertiles by normalized Step 1 entropy and exhaustively chooses depths that do not decrease from the easy group to the hard group under the corresponding budget on the mean number of steps of the primary policy for early exit. Table~\ref{tab:difficulty} assigns depths $1/3/5$ on SSC and achieves 81.12\% with a mean of 3.06 steps, compared with 81.42\% at 3.28 steps for confidence stopping. On GSC, the coarse $1/1/2$ allocation achieves 95.61\% with a mean of 1.35 steps, whereas confidence stopping reaches 95.99\% with a slightly lower mean of 1.32 steps. Step 1 entropy nevertheless correlates with the depth at which the confidence policy exits ($\rho=0.8811$ on SSC and $0.6031$ on GSC). Initial uncertainty is therefore informative, but an assignment to three bins discards sample-level variation used by the confidence policy.

\begin{figure}[H]
  \centering
  \includegraphics[width=\linewidth]{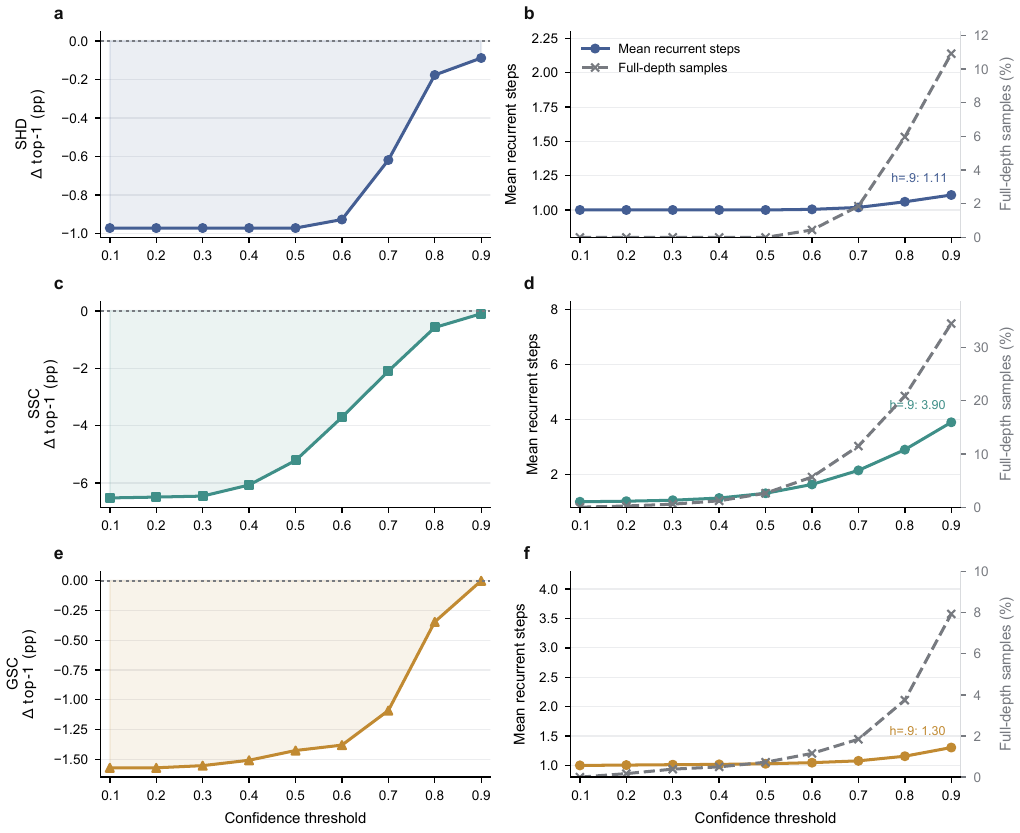}
\caption{Complete confidence-threshold scan. Accuracy is relative to full adaptive inference; the right panels show the steps actually evaluated under each threshold and the fraction reaching full task depth. Reported thresholds are selected on calibration rather than from these evaluation curves.}
  \label{fig:confidence_compute}
\end{figure}

The fixed-depth value used in the main paper linearly interpolates accuracy after a prefix of steps from the same frozen model selected for each task between adjacent integer prefix lengths at the dynamic policy's mean number of evaluated steps; it is not drawn from independently trained depth-search models. It is a descriptive control for comparison at the same average compute rather than a deployable fixed-depth system. The policy advantage on SSC and GSC indicates that confidence assigns recurrent steps more effectively than uniform truncation at the same average per-sample compute.

Fig.~\ref{fig:confidence_compute} exposes the full, uncalibrated threshold frontier. At $h=0.9$, SHD evaluates an average of 1.11 steps out of 2 and remains 0.09 points below full inference; SSC evaluates an average of 3.90 steps out of 8, with 34.5\% of samples reaching the full task depth, and remains 0.10 points below full inference; GSC evaluates an average of 1.30 steps out of 4, with only 7.9\% reaching full depth, while matching full accuracy. These evaluation curves explain why a single numerical threshold is not transferable across tasks. Reported policies instead use validation calibration and the tolerance defined in Section~\ref{app:reproducibility}.

\section{Recurrent Dynamics and Temporal Evidence}

\subsection{Evolution of the binary spike state and prediction trajectories}
\label{app:trajectories}

For Fig.~\ref{fig:reasoning}, trajectories initially wrong but ending correctly account for 34/2,264 SHD samples (1.50\%), 1,933/20,382 SSC samples (9.48\%), and 265/11,005 GSC samples (2.41\%). Among these corrected samples, delayed corrections account for 35.08\% on SSC and 24.53\% on GSC, while oscillatory corrections account for 11.28\% (218 samples) and 6.04\% (16 samples), respectively. SHD uses only two recurrent steps, so repeated changes in correctness cannot occur. Conversely, 0.53\%, 3.06\%, and 0.84\% of SHD, SSC, and GSC samples begin correctly but end incorrectly. These statistics follow successive predictions of the same trained model. For example, the first step of the SSC model trained at $D=8$ achieves 75.22\%, whereas a separate model trained at $D=1$ achieves 79.98\%.

\begin{figure}[!htbp]
  \centering
  \includegraphics[width=\linewidth]{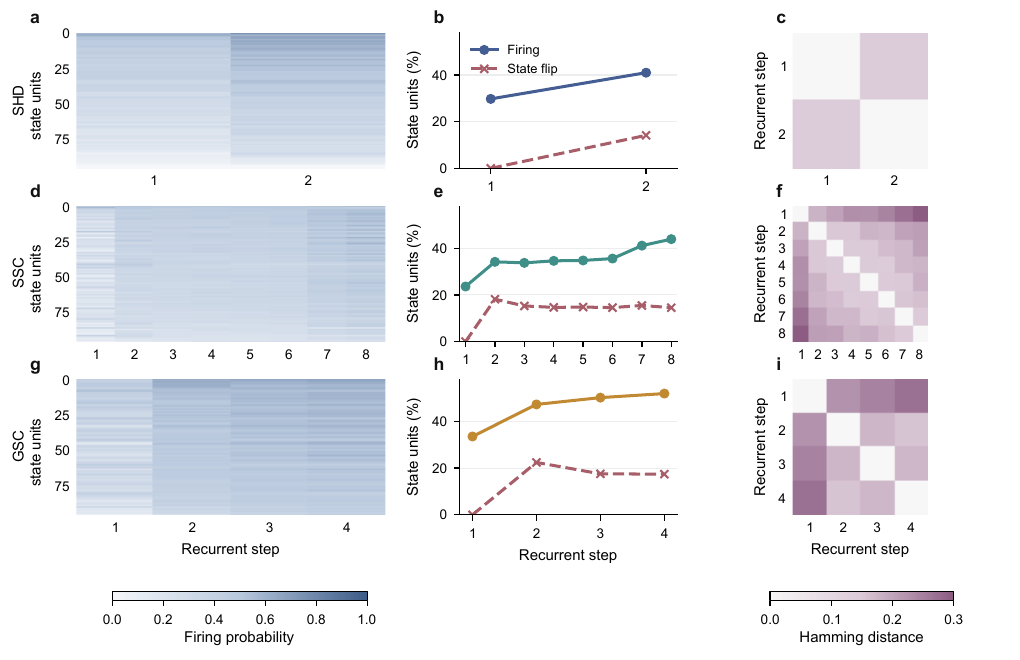}
\caption{Dynamics of the recurrent binary spike state. Firing probabilities for each neuron and step, population firing and flip rates, and pairwise Hamming distances show that the binary state vectors evolve across steps. Neurons are independently sorted per task for visualization.}
  \label{fig:state_dynamics}
\end{figure}

Fig.~\ref{fig:state_dynamics} rules out a static-state interpretation of the shared CTU. SHD state firing rate rises from 29.76\% to 41.00\% over two steps, with 14.14\% of state bits changing during the update. SSC firing rises from 23.60\% to 43.89\%; after the larger change between Step 1 and Step 2 (18.15\%), consecutive flip rates remain between 14.51\% and 15.47\%. GSC firing increases from 33.57\% to 51.95\%, while its flip rate decreases from 22.45\% at Step 2 to 17.42\% at Step 4. The nonzero pairwise Hamming distances further show that the binary state vectors differ across steps, even when the top-1 prediction is unchanged.

The state evolution also separates rapid initial correction from continued recurrent processing. From Step 1 to 2, top-1 accuracy rises by 0.97 points on SHD, 4.05 on SSC, and 1.23 on GSC, coinciding with each task's largest early change in the binary state. SSC then gains another 2.37 points by Step 8 even though its later state-flip rate stays near 15\%, while GSC gains another 0.35 points by Step 4 as its flip rate decreases. The pairwise Hamming panels show that nonadjacent steps become increasingly separated, particularly on SSC, so the later updates are neither identical replay nor a simple two-state oscillation. Their diminishing accuracy return despite persistent state motion motivates aggregation and stopping based on prediction evidence rather than state change alone.

\begin{figure}[!htbp]
  \centering
  \includegraphics[width=\linewidth]{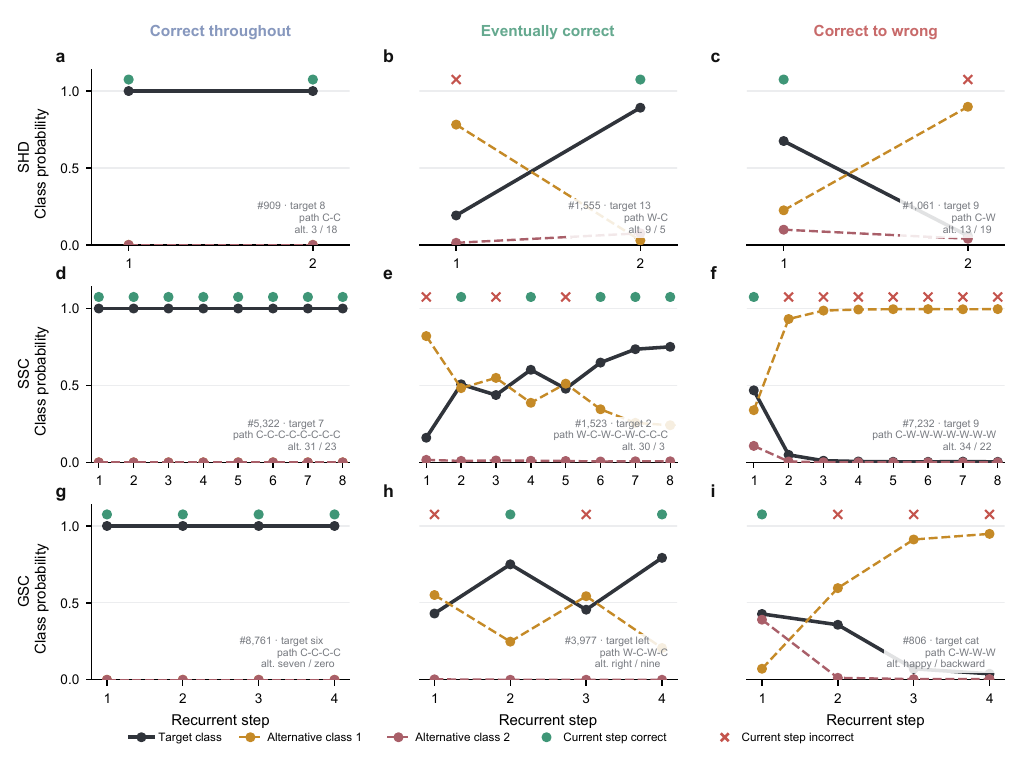}
\caption{Representative trajectories for samples that remain correct, end correctly after initially being wrong, and change from correct to wrong. Examples illustrate path shape only; prevalence claims are based on the counts reported in Fig.~\ref{fig:reasoning}.}
  \label{fig:trajectory_examples}
\end{figure}

Fig.~\ref{fig:trajectory_examples} connects those changes in the population state to class evidence. The selected SSC path follows wrong--correct--wrong--correct--wrong--correct--correct--correct while the target-class probability rises from 0.160 to 0.751; the adaptive aggregator assigns 66.9\% of its weight to the final two steps, whose predictions remain stable. The GSC example follows wrong--correct--wrong--correct and assigns 50.6\% to the recovered final step. Examples that change from correct to wrong provide the complementary failure mode: later recurrent updates can reinforce a competing class after an initially correct decision. These are deterministic representatives of path morphology, not frequency evidence; Analysis of prediction trajectories finds at least one oscillatory path in 218 of the 1,933 samples that end correctly on SSC and 16 of the 265 samples that end correctly on GSC.

\subsection{Temporal attention across steps}
\label{app:attention}

\begin{table}[H]
\centering
\caption{SSC attention migration and temporal input-occlusion results on the independent test set.}
\label{tab:attention_full}
\scriptsize
\setlength{\tabcolsep}{3.4pt}
\begin{minipage}[t]{0.47\linewidth}
\centering
\textbf{(a) Attention migration}\\[2pt]
\begin{tabular}{@{}lcc@{}}
\toprule
Trajectory & Samples & JS div. ($\times10^{-2}$) \\
\midrule
Correct throughout & 14,332 & 2.34 \\
Wrong to correct & 1,933 & 2.59 \\
Correct to wrong & 624 & 2.60 \\
\bottomrule
\end{tabular}
\end{minipage}\hfill
\begin{minipage}[t]{0.47\linewidth}
\centering
\textbf{(b) Temporal occlusion}\\[2pt]
\begin{tabular}{@{}lcc@{}}
\toprule
Occlusion & Accuracy (\%) & Drop (pp) \\
\midrule
Step 1 top 10\% & 63.29 & 18.45 \\
Step 8 top 10\% & 64.09 & 17.65 \\
Random 10\% & 77.19 & 4.55 \\
Step 8 bottom 10\% & 81.16 & 0.57 \\
\bottomrule
\end{tabular}
\end{minipage}
\end{table}

We capture all eight reader-head attention distributions without changing logits. We measure attention migration using the Jensen--Shannon divergence between attention distributions at consecutive steps. The target margin is the target-class logit minus the largest non-target logit. Table~\ref{tab:attention_full} shows that trajectories with a correction and trajectories that degrade both show greater attention migration than trajectories that remain correct: consecutive JS divergence rises from 0.02339 to 0.02588 and 0.02602, respectively. At the transition level, updates from wrong to correct have a JS divergence of 0.02785 and a flip rate in the binary state of 0.19142, compared with 0.02399 and 0.15217 when correctness is unchanged. Migration correlates with change in the binary state ($\rho=0.3818$) but only weakly with immediate target margin change ($\rho=0.0728$). Attention motion is therefore a marker of active belief revision, not evidence that the revision is necessarily beneficial.

Fig.~\ref{fig:attention_examples} illustrates this distinction at the sample level. The example that remains correct maintains concentrated attention around its dominant event region and maintains high certainty. The corrected example reallocates mass across early and middle event segments before stabilizing on the target. In contrast, the degraded example broadens and shifts attention away from the initially dominant region; certainty subsequently recovers despite the wrong class. Recurrent re-reading can thus preserve, repair, or amplify a mistaken belief.

The three paths further separate certainty from correctness. In the corrected example, the prediction changes from wrong to correct at Step 2 as attention is redistributed over the early event region and a secondary region near the utterance middle; certainty then rises from roughly 0.65 toward one. In the degraded example, a sharp initial read near 400 ms supports the correct Step 1 decision, but later attention broadens, the prediction becomes wrong, and certainty nevertheless recovers to approximately 0.8--0.9. The stable sample, meanwhile, assigns most aggregation mass to later steps even though its class decision never changes. These contrasts explain why the aggregator conditions on recurrent state and step position in addition to certainty, and why certainty alone is insufficient as an interpretation of prediction quality.

\begin{figure}[H]
  \centering
  \includegraphics[width=\linewidth]{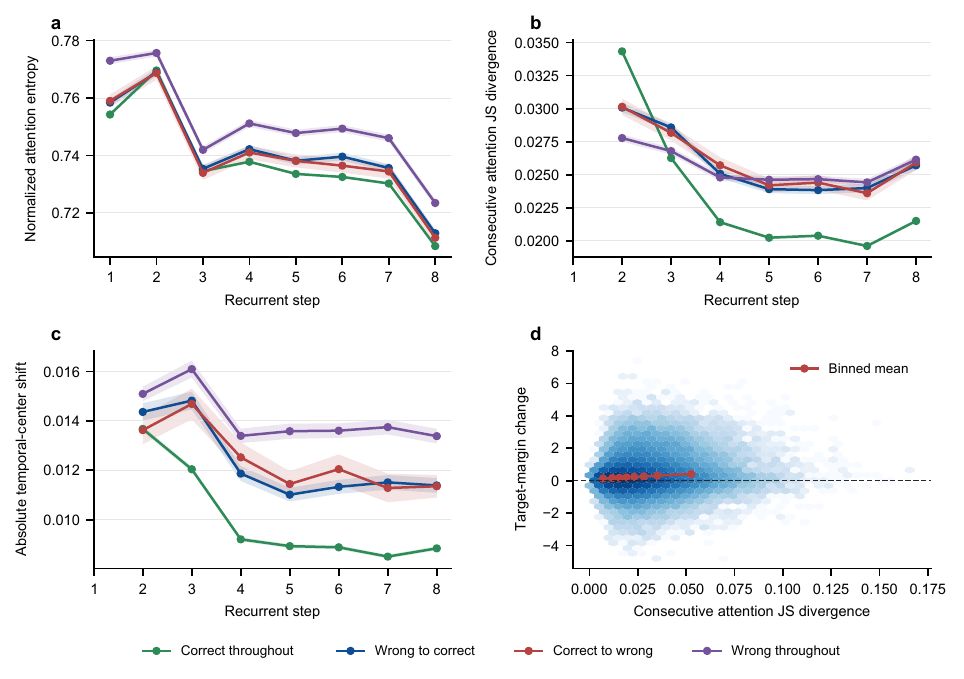}
  \caption{Population attention entropy, consecutive Jensen--Shannon divergence, center shift, and relation between migration and target margin change. Shading denotes standard error over test samples.}
  \label{fig:attention_population}
\end{figure}

Fig.~\ref{fig:attention_population} confirms that migration is not confined to selected examples. Among 20,382 SSC test samples, 14,332 remain correct throughout, 1,933 are corrected, 624 begin correctly but end incorrectly, and the remaining 3,493 never reach the correct class. The corrected and degraded groups both exceed the JS level for trajectories that remain correct. The bootstrap intervals for corrected and stable trajectories do not overlap. However, the weak $\rho=0.0728$ relation to immediate margin change shows that the magnitude of attention movement alone does not determine the direction of the class update.

\begin{figure}[H]
  \centering
  \includegraphics[width=0.78\linewidth]{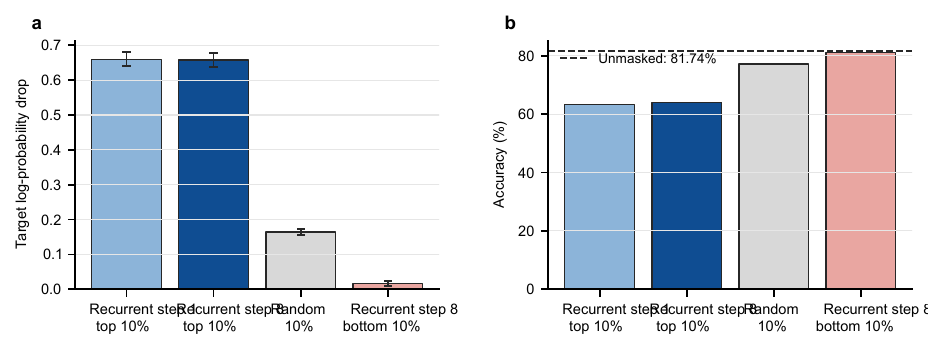}
\caption{Temporal occlusion of the input. Masking the highest-attention bins is substantially more damaging than masking matched random bins or the lowest-attention bins; the dashed line is unmasked accuracy.}
  \label{fig:attention_occlusion}
\end{figure}

Table~\ref{tab:attention_full}b and Fig.~\ref{fig:attention_occlusion} test whether the attended input regions are decision-relevant rather than relying on attention weights alone. Masking the top 10\% of bins attended at the final step reduces accuracy from 81.74\% to 64.09\%, compared with 77.19\% under matched random masking and 81.16\% after bottom-bin masking. The target-log-probability drop exceeds random masking by 0.4931 (bootstrap 95\% interval $[0.4749,0.5116]$) and bottom masking by 0.6412 (bootstrap 95\% interval $[0.6209,0.6611]$). Step 1 top-bin masking is similarly damaging: the paired target-log-probability difference (masking based on the final step minus Step 1 masking) is only $-0.0023$ (bootstrap 95\% interval $[-0.0172,0.0142]$), indicating no meaningful difference in damage between the two masking conditions. Attended regions are therefore decision relevant, but later steps do not monotonically discover better temporal locations; they dynamically reweight evidence that is already informative.

\section{Operation Proxies and Energy Boundaries}
\label{app:energy}

\begin{table}[H]
\centering
\caption{Operation counts at the full depth selected for each task and 45-nm-process proxy energy at $\T=100$.}
\label{tab:energy_full}
\small
\begin{tabular}{lccccc}
\toprule
Data & $D$ & SOPs (G) & MACs (G) & Total ops (G) & Energy (mJ) \\
\midrule
SHD & 2 & 0.0014 & 0.0028 & 0.0041 & 0.0140 \\
SSC & 8 & 0.0016 & 0.0031 & 0.0047 & 0.0158 \\
GSC & 4 & 0.0006 & 0.0049 & 0.0055 & 0.0231 \\
\bottomrule
\end{tabular}
\end{table}

\begin{table}[H]
\centering
\caption{Theoretical lower bound obtained by running one step; tokenizer and memory construction remain included.}
\label{tab:energy_lower}
\small
\begin{tabular}{lccccc}
\toprule
Data & Task-selected full $D$ & Accuracy (\%) & SOPs (G) & Energy (mJ) & Energy reduction (\%) \\
\midrule
SHD & 2 & 95.63 & 0.0013 & 0.0139 & 0.76\% \\
SSC & 8 & 75.22 & 0.0015 & 0.0150 & 4.90\% \\
GSC & 4 & 94.38 & 0.0005 & 0.0227 & 1.50\% \\
\bottomrule
\end{tabular}
\end{table}

For a spike-driven layer $l$, we estimate $\mathrm{SOP}_l=\mathrm{FR}_{l-1}\mathrm{FLOP}_l$ from its measured input firing rate; operations on continuous paths remain MACs. Proxy energy uses $0.9$ pJ per accumulation operation and $4.6$ pJ per multiply--accumulate. Table~\ref{tab:energy_full} shows that GSC has the lowest SOP count (0.0006 G) but the highest proxy energy (0.0231 mJ), because its 0.0049 G continuous MAC path dominates both total operations and the higher per-operation energy coefficient. SHD and SSC have more SOPs but smaller MAC totals, yielding lower proxy energy. Sparse-operation count alone is therefore insufficient for comparing hybrid models.

Table~\ref{tab:energy_lower} gives the theoretical lower bound obtained by running one step while retaining tokenizer and memory construction. Relative to depth selected for each task, it lowers proxy energy by only 0.76\% on SHD, 4.90\% on SSC, and 1.50\% on GSC, while reducing accuracy by 0.97, 6.52, and 1.58 points. The limited energy change shows that most estimated cost is paid before recurrent depth can be shortened; SSC has the largest recurrent share and consequently the largest energy reduction from running one step, but also the largest accuracy loss. Neither table includes memory movement, nonlinear functions, normalization, spike-neuron updates, control, or device utilization, and neither replaces measurement on a specified hardware platform. For spike-form inputs (SHD, SSC), SpikCommander's convention treats all operations as ACs~\citep{wang2026spikcommander}; we instead count continuous residual paths as MACs, so part of the SHD energy difference (0.0140 versus 0.0039 mJ) reflects accounting conventions rather than computational work. As in recent spiking speech work, theoretical 45-nm-process proxy energy is a common field-standard estimate in the absence of device access; deployment and measurement on neuromorphic hardware remain future work.

\end{document}